\documentclass[12pt]{article}

\usepackage{scicite}
\usepackage{graphicx}
\usepackage[superscript,biblabel]{cite}

\usepackage{times}

\newenvironment{sciabstract}{%
\begin{quote}}
{\end{quote}}

\newcounter{lastnote}

\begin{document}

\begin{center}
{\LARGE Deep learning from wristband sensor data: towards wearable, non-invasive seizure forecasting}\\
\vspace{0.8cm}
{\large Christian Meisel$^{1,2\ast}$}\\
{\large Rima El Atrache $^{1}$}\\
{\large Michele Jackson $^{1}$}\\
{\large Sarah Schubach $^{1}$}\\
{\large Claire Ufongene $^{1}$}\\
{\large Tobias Loddenkemper$^{1}$}\\
\end{center}
{\normalsize$^{1}$ Boston Children's Hospital, Boston, USA}\\
{\normalsize$^{2}$ Department of Neurology, University Clinic Carl Gustav Carus, Dresden, Germany}\\

{\normalsize$^{\ast}$ corresponding author email address: christian@meisel.de}\\
\\


\date{}



\baselineskip24pt



\pagebreak

\begin{sciabstract}
Seizure forecasting may provide patients with timely warnings to adapt their daily activities and help clinicians deliver more objective, personalized treatments. 
While recent work has convincingly demonstrated that seizure risk assessment is possible, these early approaches relied largely on complex, often invasive setups including intracranial electrocorticography, implanted devices and multi-channel EEG, which limits translation of these methods to broad clinical application.
To facilitate broader adaptation of seizure forecasting in clinical practice, non-invasive, easily applicable techniques that reliably assess seizure risk, in combination with clinical information, are crucial. 
Wristbands that continuously record physiological parameters, including electrodermal activity, body temperature, blood volume pressure and actigraphy, may afford monitoring of autonomous nervous system function and movement relevant for such a task, hence minimizing potential complications associated with invasive monitoring, and avoiding stigma associated with bulky external monitoring devices on the head.
Here, we use deep learning to analyze long-term, multi-modal wristband sensor data from 50 patients with epilepsy (total duration $>$1400 hours) to assess its capability to distinguish pre- from interictal states. Prediction performance is assessed using area under the receiver operating charateristic (AUC) and improvement over chance (IoC) based on F1 scores. Using one- and two-dimensional convolutional neural networks, we identified better-than-chance predictability in out-of-sample test data in 60\% of the patients in leave-one-out and 43\% of patients in pseudo-prospective approaches. 
These results provide a step towards developing easier to apply, non-invasive methods for seizure risk assessments in patients with epilepsy. 
\end{sciabstract}

\section*{Introduction}

Reliable methods to assess seizure risk could alleviate a major burden for epilepsy patients by providing timely warning or relief when seizure risk is high or low. From a clinician perspective, robust seizure risk assessments are desirable because of their ability to improve treatment by optimizing dosing and timing of antiepileptic drug regimen by objective, personalized standards, as well as by potentially enabling timely interventions to avert impending seizures \cite{Dumanis2017}.

Following initial attempts \cite{Mormann2007}, there has been a recent surge of studies demonstrating the possibility of accurate seizure forecasting \cite{Freestone2017, Kaggle2016, Mormann2016, Kuhlmann2018}. To this end, most studies have utilized either electrocorticography (ECoG) or scalp electroencephalography (EEG) as well as, to a lesser extent, electrocardiography (ECG), and have demonstrated that robust differentiation between preictal and interictal states is possible with a performance better than chance \cite{Cook2013, Lehnertz2016, Fujiwara2016, Kerem2005}. In order to make seizure risk assessments available for broader clinical use, however, methods that build on non-invasive, easily recordable data streams are desirable \cite{SchuBo2010}. Peripheral signals recorded using wearable devices, such as wristbands, are particularly interesting in this respect since these signals permit continuous, non-invasive recording of several physiological parameters. At the same time, the compact design may limit the risk of stigmatization, are easy to apply, and may altogether increase patient adherence relevant for long-term ambulatory use. 

Continuous and simultaneous monitoring of a range of physiological parameters, such as electrodermal activity, body temperature, blood volume pressure and actigraphy, using wristbands is becoming increasingly available and permits close monitoring of autonomous nervous system function and movement \cite{Poh2010}, and may assist in the detection of generalized tonic-clonic seizures \cite{Poh2012}. 

Similar autonomous system measures may also provide information on detection of preictal patterns or states. Deep learning has been shown to exhibit strong classification performance from complex feature sets \cite{Schmidhuber2015}. It therefore constitutes a promising technique to differentiate pre- from interictal states from complex, multi-modal wristband data. While more traditional machine learning approaches rely on hand-designed feature sets, deep learning uses multiple layers of connections to perform classification tasks without the need of feature designing, which may be an advantage in relatively under-explored, multi-modal datasets, such as data from wristworn devices.

In this work, we use convolutional neural networks (CNNs) on a unique dataset comprised of multi-modal wristband sensor data recorded from patients with epilepsy during multi-day in-hospital monitoring. Our aim is to evaluate the utility of this novel data and methods approach in its ability to differentiate pre- from interictal states as a prerequisite for future, non-invasive seizure forecasting methods. 

\section*{Materials and Methods}
\subsection*{Data recording and preprocessing}
We recruited patients with epilepsy admitted to the long-term video-EEG monitoring (LTM) unit and placed a biosensor wristband (E4, Empatica \cite{Poh2010}) on either left or right wrist or ankle for long-term recording. For the purpose of this study we considered all patients with wristband recordings from 02/2015 until 11/2018. For the purpose of this study, we considered data from one wristband per patient only. 
When patient recording involved multiple wristbands (e.g. from wrist and ankle), we selected the data from the biosensor wristband with the longest total recording time for further analysis. 

\begin{figure}[htbp]
\centering
\includegraphics[width=0.75\textwidth]{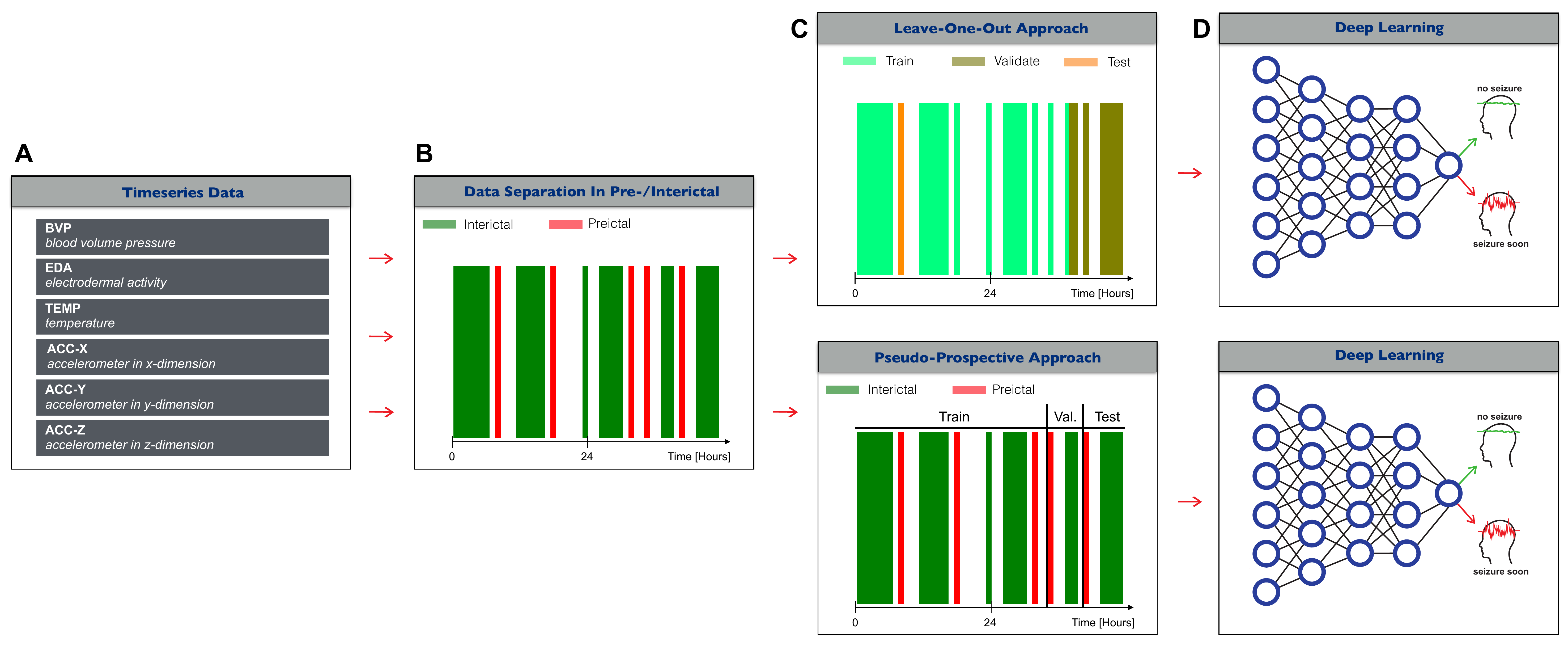}
\caption{\label{fig_1}
Schematic outline of data processing in order to predict pre- and interictal states. A, continuous recording of timeseries data. B, data separation into preictal and interictal segments. C, separation into training, validation and test data for each approach, leave-one-out (example shown for first seizure) and pseudo-prospective. D, training, validation and out-of-sample testing using convolutional neural networks.}
\end{figure}

\begin{figure}[htbp]
\centering
\includegraphics[width=0.75\textwidth]{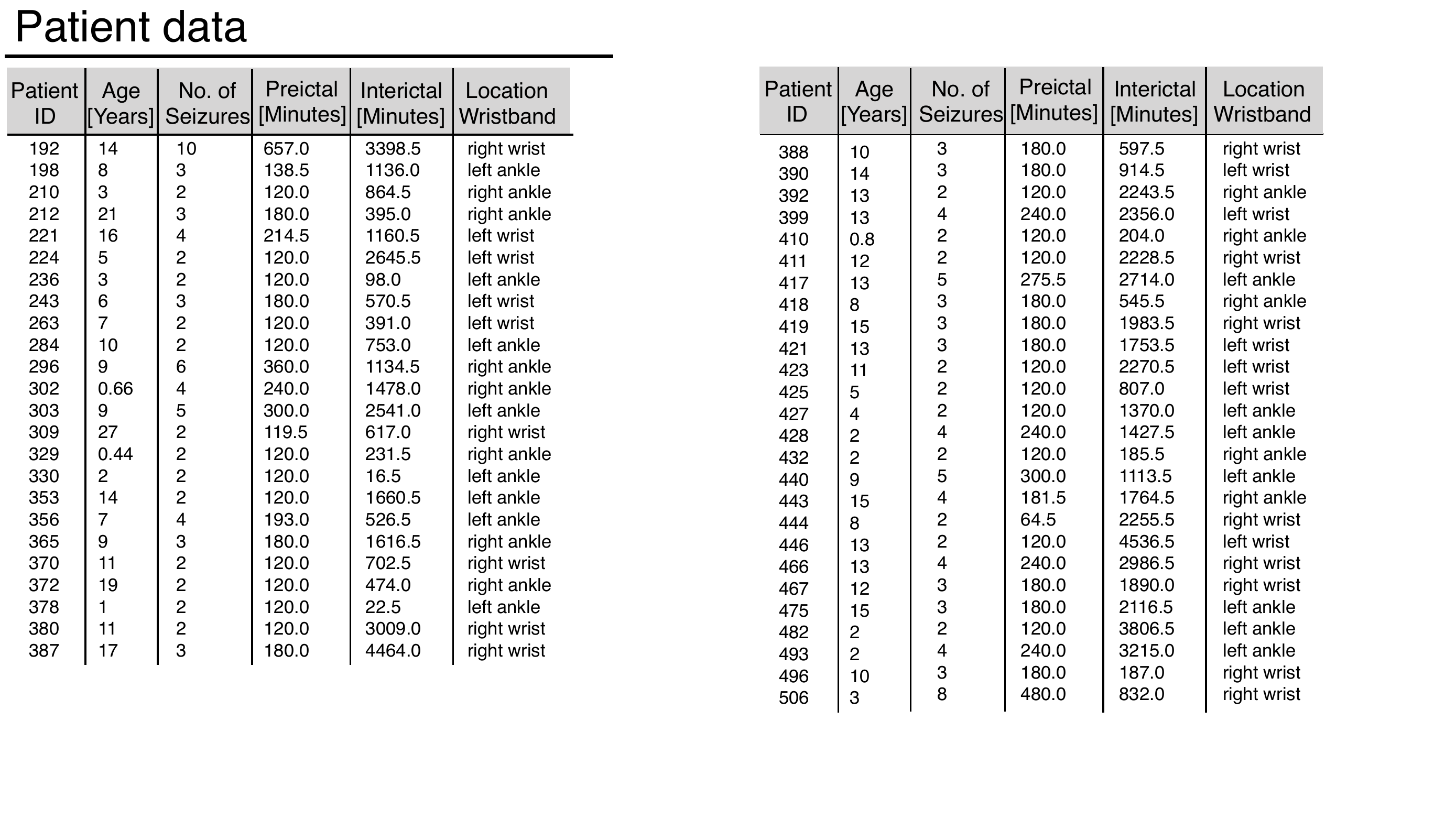}
\caption{\label{table_1}
Summary of patient characteristics.}
\end{figure}

A prerequisite for seizure risk assessment is the reliable distinction between pre- and interictal states. We aimed to differentiate pre- from interictal states based on 30-second wristband recordings composed of six sensor data streams (electrodermal activity (EDA), accelerometer data in three dimensions, blood volume pressure (BVP), and temperature (TEMP); Fig. \ref{fig_1} A). We considered a 30-second data segment as preictal if it occurred between 61 minutes and one minute prior to a seizure, thus leaving a one minute buffer prior to seizure onset (Fig. \ref{fig_1} B, red). Electrographic seizure onset was determined using video and EEG recordings. We analyzed all epileptic seizure types occurring in a patient, which included subclinical seizures, focal, primary and secondary tonic-clonic, myoclonic, clonic, tonic, atonic seizures and epileptic spasms (Figure \ref{table_1}). Only seizures that occurred two hours or more from a preceding seizure were included to limit our analysis on lead seizures. 30-second data segments were classified as interictal if they occurred two hours or more from any seizure (Fig. \ref{fig_1} B, green). To allow stable recording conditions, we removed data from the first and last hour of each recording. 

\subsection*{Separation in training, validation and test data}
To evaluate classification performance on out of-sample test data, we used two separate approaches, a leave-one-out approach and a pseudo-prospective approach.
\subsubsection*{Leave-one-out approach}
A leave-one-out cross-validation approach has the advantage that all seizures can be used for training and validation. Here, we used a leave-one-out approach for each seizure. Similar to \cite{Truong2018}, if a subject had $N$ seizures, $(N - 1)$ seizures were used for training/validation (75\%/25\% split in temporal order), and the remaining seizure was used for testing. This was done $N$ times, so that all seizures were used for testing once. Interictal segments were randomly split into $N$ parts, where $(N - 1)$ parts were used for training/validation (75\%/25\% split in temporal order) and the remaining part was used for testing. Fig. \ref{fig_1} C (top) illustrates the leave-one-out separation using the first preictal (seizure) segment. The leave-one-out approach consequently required patients with at least two seizures which reduced the analysis to 50 patients (Figure \ref{table_1}). 

\subsubsection*{Pseudo-prospective approach}
For seizure prediction under real-world conditions, the risk for seizures at a given time needs to be assessed using algorithms trained and validated on past data. A pseudo-prospective evaluation method that only evaluates an algorithm using out-of-sample data recorded at a later point in time than the data used for algorithm training and validation matches these conditions most closely. A pseudo-prospective evaluation setup benefits from large data sets with sufficient data for training and pseudo-prospective validation and testing. For this purpose, we separated data into non-overlapping, consecutive training, validation and test data, where validation and test data contained preictal segments from one seizure each and training data the remaining, preceding preictal and interictal periods (Fig. \ref{fig_1} C, bottom). The exact cut-off between validation and test data was chosen so that both contained an equal amount of interictal data. This procedure guaranteed that training and validation data always chronologically preceded test data and that training, validation and test data were chronologically separated. This constraint on data, which required patients to have had at least three seizures, reduced the analysis to 21 patients whose data could be separated for such a pseudo-prospective approach.

\subsection*{Neural networks and training}
\begin{figure}[htbp]
\centering
\includegraphics[width=0.75\textwidth]{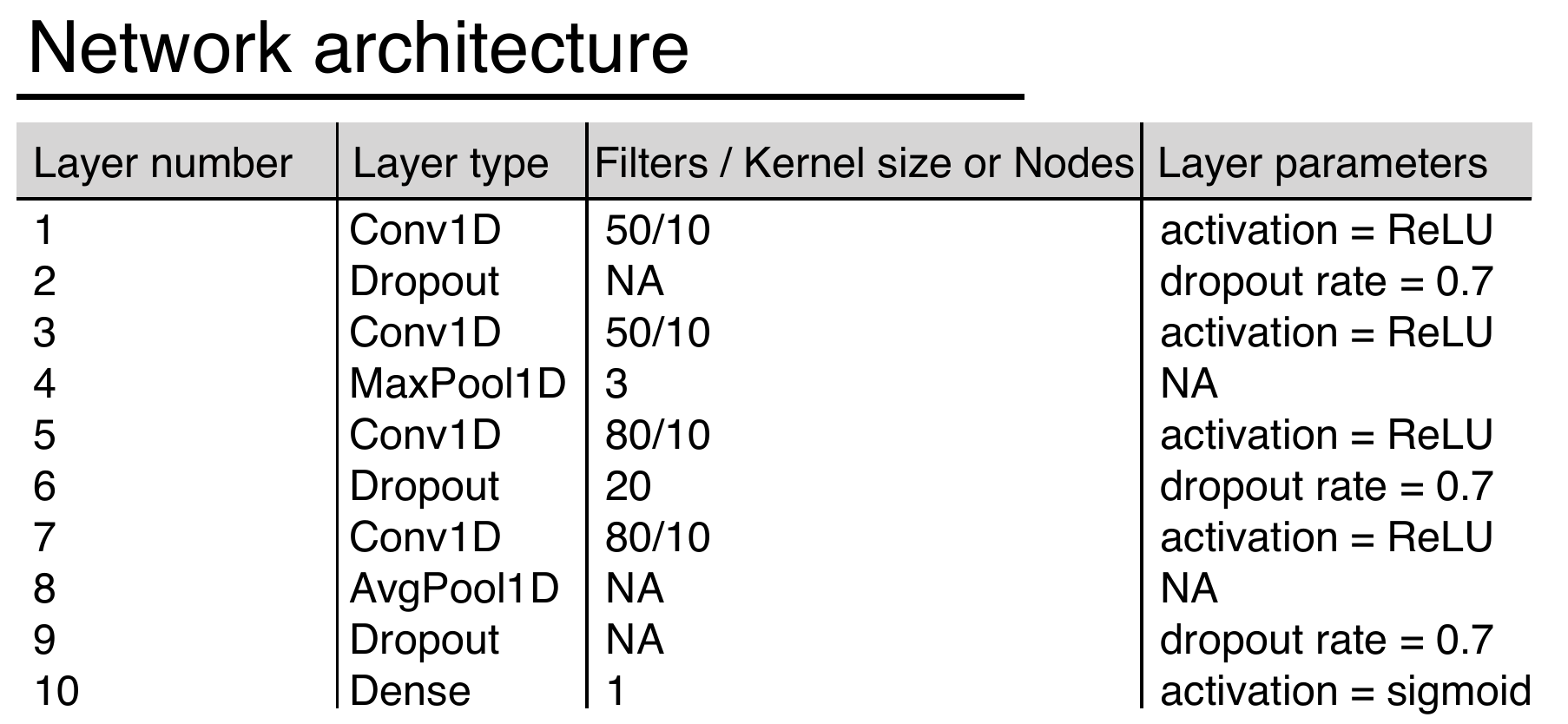}
\caption{\label{table_2}
One-dimensional convolutional (1DConv) network topology used. Learning rate was set to 0.0001.}
\end{figure}

We primarily used one-dimensional convolutional neural networks (CNNs), as they have been shown to provide robust classification performance based on multi-dimensional timeseries data \cite{Schmidhuber2015}. To use the wristband sensor data in a one-dimensional CNN (1DConv), data was down-sampled to 4 Hz for all sensors in order to provide the same vector length for each 30-second segment. Figure \ref{table_2} shows a summary of the CNN parameters used. 

We also compared our results to a two-dimensional convolutional neural network (2DConv) that used the power spectral density (PSD, FFT routine; 30-second, non-overlapping Hanning windows) of signals sampled at 64 Hz (if originally sampled below 64 Hz, signals were upsampled by repeating values). PSD data were then reshaped in 31 by 31 images of depth 6 for use in neural networks. The use of 2DConv networks using PSD was motivated by encouraging results of similar approaches using EEG and ECoG data \cite{Truong2018, Kiral-Kornek2018}.

We used balanced learning to handle imbalanced training sets, repeating either pre- or interictal segments in the training set until a 50-50 preictal-interictal balance was achieved.
All networks were trained for 1000 epochs. To limit CNNs from overfitting, we kept the CNN architecture simple and shallow.

\subsection*{Performance metrics and statistical tests}
Performance was assessed on out-of-sample test data using the area under the receiver operating curve (AUC) and improvement over chance based on F1 scores (IoC). AUC is commonly used to assess performance in classification problems. AUC scores were obtained for ten network executions (pseudo-prospective approach) or all folds (leave-one-out approach, five network executions each). A one-sample t-test was used to compare AUC-values against chance (i.e., AUC=0.5).

The F1 score is the harmonic mean between precision and recall and, as a single metric, evaluates how well a classification algorithm performs on the minority class. The use of F1 scores as an additional method was motivated by the observation that F1 scores may provide more reliable performance estimates in skewed, imbalanced data sets \cite{Chawla2005,Jeni2013}, such as in classification problems in epilepsy where interictal data typically outnumber preictal data. $F1_{Test}$ scores were obtained for ten network executions (pseudo-prospective approach) or all folds (leave-one-out approach, five network executions each). To compare our algorithms to a chance predictor, the F1 score was obtained for random predictions (i.e., randomly shuffled test set lablels, which maintains the numbers of pre- and interictal classifications but unlinks any correlation to the data) averaged across 1000 randomizations (pseudo-prospective approach; 1000 randomization per fold for the leave-one-out approach) resulting in a mean $F1_{Chance}$ score. Improvement over chance (IoC) was then defined as $IoC=\frac{F1_{Test} - F1_{Chance}}{1-F1_{Chance}}$ and used as the main performance metric in this article. A one-sample t-test was used to compare IoC-values against chance (i.e., IoC=0). IoC has been repeatedly used in epilepsy forecasting research as a meaningful metric to characterize algorithm performance \cite{Mormann2007, Cook2013, Kiral-Kornek2018}.

\section*{Results}
We assessed the utility of CNNs in distinguishing pre- from interictal states using long-term, multi-modal wristband sensor data \cite{Poh2010} obtained during epilepsy monitoring. Neural network based assessments were done for each patient individually and for two approaches, a leave-one-out cross-validation approach and a pseudo-prospective approach (Fig. \ref{fig_1}).

\subsubsection*{Leave-one-out approach}
The Leave-one-out approach required patient datasets to contain at least two seizures. With this constraint, we analyzed a total of 50 patients with a total of 157 seizures and a recording time of 1425.92 hours (1270.18 hours interictal, 155.74 hours preictal; Figure \ref{table_1}). 

\begin{figure}[htbp]
\centering
\includegraphics[width=0.75\textwidth]{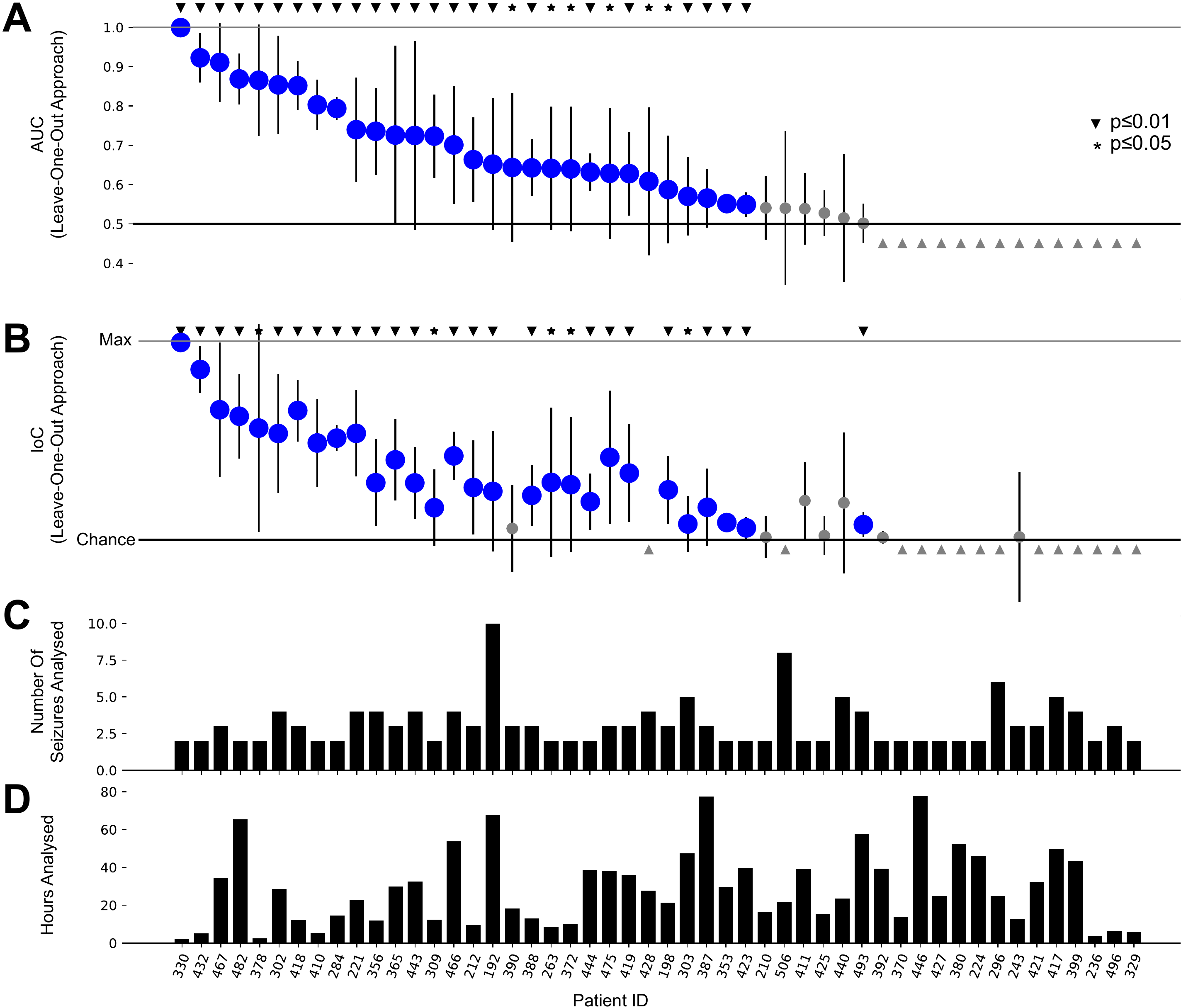}
\caption{\label{fig_2}
Prediction performance for the leave-one-out approach using a one-dimensional convolutional network (1DConv, n=50 patients). A, area under the receiver operating curve (AUC;  blue markers indicate a performance significantly better than chance). B, improvement over chance (IoC). C, seizure count. D, total duration of data analyzed.}
\end{figure}

Each fold of the leave-one-out approach was executed five times, and average results with standard deviations were reported for each patient. Area under the receiver operating curve (AUC) values and improvement over chance (IoC) values using the 1DConv network are depicted along with seizure number and data duration for all patients in figure \ref{fig_2}. Overall, a prediction better than chance was obtained for more than half of the patients (Fig. \ref{fig_2} A, B, blue markers; AUC significant for 30/50 patients, $AUC=0.714\pm0.121$; IoC significant for 29/50 patients, $IoC=0.377\pm0.229$). In comparison, the 2DConv network did not exhibit an overall better performance (AUC significant for 26/50 patients, $AUC=0.595\pm0.069$; IoC significant for 29/50 patients, $IoC=0.164\pm0.131$).


\subsubsection*{Pseudo-prospective approach}
\begin{figure}[htbp]
\centering
\includegraphics[width=0.75\textwidth]{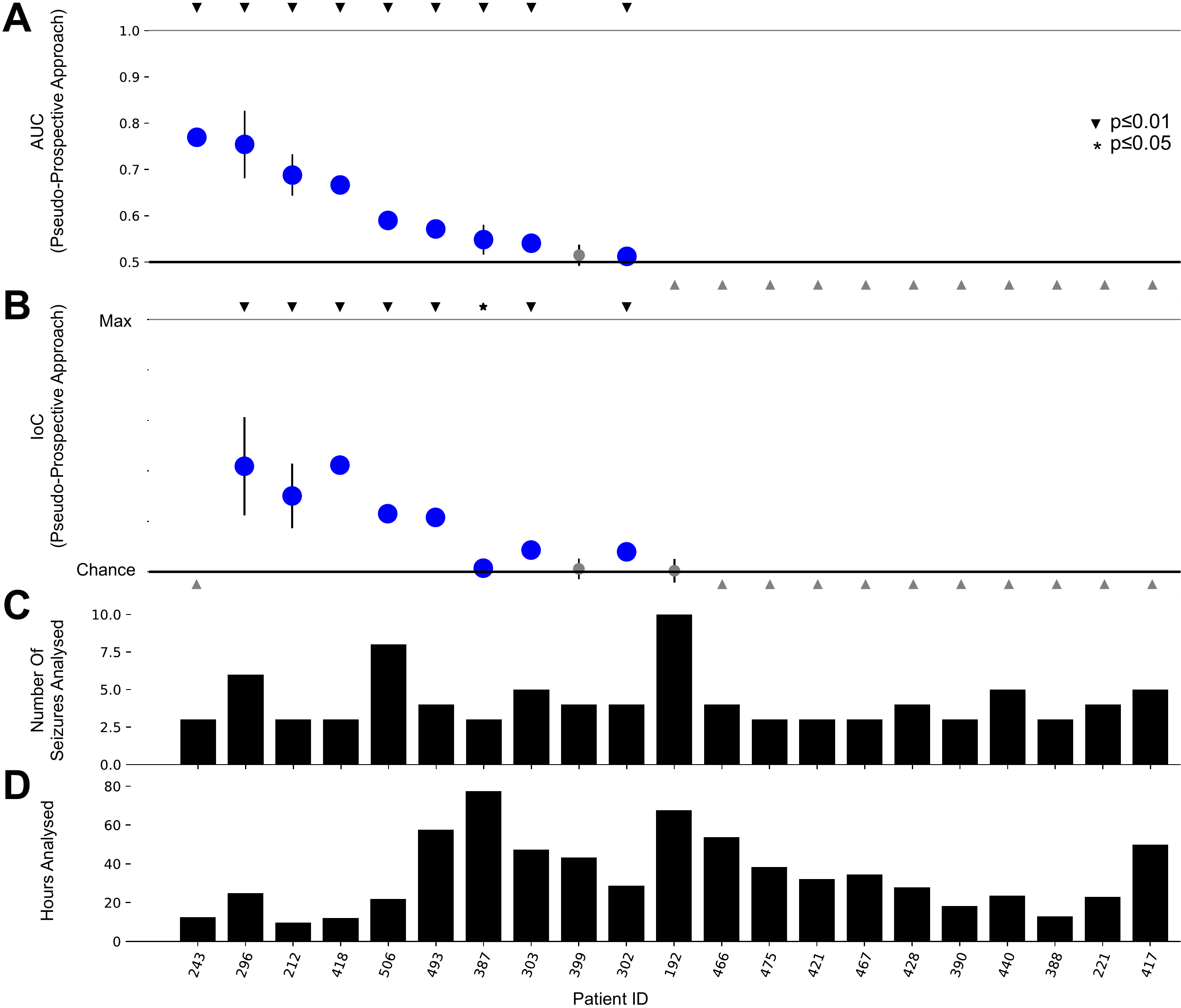}
\caption{\label{fig_3}
Prediction performance for the pseudo-prospective approach using a one-dimensional convolutional network (1DConv, n=21 patients). A, area under the receiver operating curve (AUC; blue markers indicate a performance significantly better than chance). B, improvement over chance (IoC). C, seizure count. D, total duration of data analyzed.}
\end{figure}

Although our data here is comparably short, we attempted a pseudo-prospective evaluation to assess our approach under more realistic conditions. Requiring validation and test data to be chronologically following training data with at least one seizure each reduced our analysis to 21 patients. Networks were executed ten times, and average results with standard deviations were reported (Fig. \ref{fig_3} A, B, blue markers; AUC significant for 9/21 patients, $AUC=0.627\pm0.090$; IoC significant for 8/21 patients, $IoC=0.221\pm0.144$). The 2DConv network again did not exhibit an overall better performance (AUC significant for 9/21 patients, $AUC=0.572\pm0.062$; IoC significant for 13/21 patients, $IoC=0.094\pm0.061$).


\section*{Discussion}
Here we assessed the utility of physiological sensor data recorded from a wristband to estimate seizure risk. The ability to robustly differentiate preictal from an interictal states is a necessary prerequisite for any seizure forecasting or seizure risk assessment \cite{Mormann2007}. For this classification process, we used convolutional neural networks which have demonstrated outstanding performance in classification tasks in several domains \cite{Schmidhuber2015}, more recently also including epilepsy \cite{Truong2018, Kiral-Kornek2018}. Our work is motivated by the potential benefits for patients and clinicians from a robust seizure gauge. Forecasting seizures would provide patients with timely warning to adapt daily activities and allow clinicians to titrate therapies and develop novel interventions that potentially could prevent impending seizures \cite{Dumanis2017, Baud2018}. Peripheral sensor data that can be recorded easily and non-invasively with a wristband would be desirable for such a purpose since approaches based on ECoG \cite{Cook2013} or a large number of scalp EEG channels \cite{SchuBo2010} limit broad clinical application.

Of the patients included in our analysis, about half displayed a significant improvement over chance (IoC) for both evaluation schemes, leave-one-out and pseudo-prospective approaches. On the one hand, these performance values may not appear as strong as what is reported in other recent studies where the majority of patients exhibited predictability levels better than chance \cite{Truong2018, Kiral-Kornek2018}. On the other hand, these results suggest that seizure forecasting might also be feasible with relatively short, noisy, multi-modal signals recorded from wristbands far away from the brain. The better-than-chance classification performance in about half of the patients was obtained despite the comparably brief duration of data, where training sometimes only involved one or two seizures, and the variability in seizure types, data duration, age and wristband location. Apart from the criteria to label pre- and interictal data, we used no other preselection or preprocessing, but instead included the data in raw format "as is" in an attempt to maximize the transferability of our approach to real-world, noisy conditions. Predictability performance across patients did not depend on wristband location, overall duration of data, or seizure type.

Seizure forecasting builds on the notion that a preictal state, during which a seizure is more likely to occur soon, can be reliably distinguished from interictal states.
To this end, most studies have focused on data recorded either from ECoG and EEG or from ECG. ECG has thus been a long-standing example that peri- and preictal changes can not only be detected within the central nervous system but are also reflected in a variety of cardiac effects \cite{VanBuren1958, Marshall1983, Smith1989}.
Cardiac activity is controlled by parasympathetic and sympathetic branches of the autonomic nervous system, with the former producing an inhibitory response and the latter producing an excitatory response on heart rate \cite{Robinson1966}. Preictal changes in brain activity that occur in or propagate to autonomic control centers may affect this autonomic balance and, consequently, affect cardiac activity during the leadup to a seizure. A recent study that compared the information content in ECoG, EEG and ECG in terms of identifying preictal states found that single-channel ECG contains a comparable amount of information to multi-channel EEG \cite{Meisel2019}, which highlights the relevance of peripheral sensors for seizure forecasting. 
Autonomous nervous system changes are captured by the wristband sensors used in this study in several ways. Electrodermal activity is known to be sensitive to sympathetic innervation. Blood volume pressure curves contain information about heart rate which is controlled by the parasympathetic and sympathetic interplay. Similarly, body temperature is known to be maintained by the autonomic nervous system.
The approach proposed in this study builds on monitoring these autonomous nervous system functions along with actigraphy, which indirectly also monitors resting periods and sleep, and therefore pioneers seizure forecasting capabilities based on such multi-modal sensor data, going beyond more traditional ECoG/EEG and ECG approaches.

In our approach, we used the same model for all patients, albeit models were trained for each patient individually. While it is possible that model hyperparameters individualized for each patient might bring about better performances, we chose to have the same model architecture across patients that could potentially be implemented "out-of-the-box" in future prospective settings.  
Previous studies using CNNs for seizure forecasting have relied on power spectral densities converted to two-dimensional images used in neural networks \cite{Truong2018, Kiral-Kornek2018}. In contrast, our approach relied on one-dimensional convolutional networks where we used raw sensor data for inputs directly, thus requiring less preprocessing. Although we did not do an extensive comparison in terms of model hyperparameters, we found that, at least for the two models chosen here, the 1DConv networks using raw data performed slightly better than the 2DConv network using the data's power spectrum. 

Our evaluation used two different approaches, each of which has certain advantages and disadvantages. Leave-one-out approaches have the advantage that all seizures can be used for training and validation to obtain an estimate of algorithm performance. This can be desirable particularly for relatively short data durations, such as ours. For this reason, leave-one-out cross-validation approaches have been used to assess seizure prediction and detection performance in epilepsy research \cite{Poh2012, Truong2018}. We here applied a leave-one-out approach to preictal intervals belonging to different seizures in order to assess the algorithmic ability to differentiate pre- from interictal states. For a forecasting algorithm to be used in real-world conditions to assess future seizure risk, classification of incoming data into pre- and interictal classes has to be performed in a prospective manner using data recorded after the training and validation periods. A pseudo-prospective assessment benefits particularly from large data sets with sufficient data for training and pseudo-prospective validation and testing \cite{Kiral-Kornek2018}.

Results need to be interpreted in the setting of data acquisition. One limitation of our study is the relative short duration of recordings which covered only a few days of continuous data per patient. Training data benefits from long periods of data where algorithms can better learn to generalize and which gives a more realistic account of seizure forecasting capabilities. However, the current dataset is unique in the sense that it contains multi-modal sensor data over several days from a relatively large number of epilepsy patients. The better-than-chance predictability in about half of the patients in this study is therefore encouraging for future, longer trials using these sensors. 
Another limitation is the absence of benchmarks to compare our approach to. While we compared our results to chance predictors in out-of-sample test data, the uniqueness and novelty of the current dataset limits more comprehensive comparison to other approaches. There is growing awareness of the benefits of creating data warehouse ecosystems that allow rigorous and continuous reevaluation and benchmarking by making data and algorithms available to many researchers \cite{Brinkmann2016}. We expect that these open-science efforts will increase the reproducibility and help benchmark and improve algorithms, such as the ones proposed in the current study, in the future. Finally, other clinically meaningful metrics, such as false alarm rate, or time between seizure warning and actual seizure, that characterize an algorithm's forecasting performance may be desirable. These metrics, however, build on top of classifications and require further post-processing. Due to the brevity of the data and the absence of other relevant benchmarks to compare our approach to, we did not consider this useful in the current context. Instead our main goal in this study was more basic, to assess whether the combination of peripheral wristband sensor data and deep learning might be able to differentiate pre- from interictal states with a better-than-chance performance. For this purpose, AUC and IoC, which have both been used in several other forecasting studies \cite{Mormann2007, Cook2013, Kuhlmann2018b, Brinkmann2016, Kiral-Kornek2018}, are valid metric which also allow future comparison and benchmarking.

Seizure forecasting is likely to bring about notable benefits for epilepsy patients and clinicians. In order to make seizure forecasting available for broad use, non-invasive, easily applicable techniques are greatly desirable. We here assessed the capability of multi-modal wristband sensor data in combination with deep learning to reliably distinguish pre- from interictal states. Our results demonstrate a better-than-random predictability in about half the patients, even when a pseudo-prospective approach on out-of-sample test data was taken. Future, more long-term studies should help to validate the utility of this approach.

\bibliography{sd_slowing}

\begin{thebibliography}{10}

\bibitem{Dumanis2017}
S.~B. Dumanis, J.~A. French, C.~Bernard, G.~A. Worrell, and B.~E. Fureman.
\newblock {{S}eizure {F}orecasting from {I}dea to {R}eality. {O}utcomes of the
  {M}y {S}eizure {G}auge {E}pilepsy {I}nnovation {I}nstitute {W}orkshop}.
\newblock {\em eNeuro}, 4(6), 2017.

\bibitem{Mormann2007}
F.~Mormann, R.~G. Andrzejak, C.~E. Elger, and K.~Lehnertz.
\newblock Seizure prediction: the long and winding road.
\newblock {\em Brain}, 130:314--333, 2007.

\bibitem{Freestone2017}
D.~R. Freestone, P.~J. Karoly, and M.~J. Cook.
\newblock {{A} forward-looking review of seizure prediction}.
\newblock {\em Curr. Opin. Neurol.}, 30(2):167--173, 04 2017.

\bibitem{Kaggle2016}
{M}elbourne {U}niversity {AES}/{M}ath{W}orks/{NIH} {S}eizure {P}rediction.
\newblock {https://www.kaggle.com/c/melbourne-university-seizure-prediction}.
\newblock Accessed: 2016-10-28.

\bibitem{Mormann2016}
F.~Mormann and R.~G. Andrzejak.
\newblock {{S}eizure prediction: making mileage on the long and winding road}.
\newblock {\em Brain}, 139(Pt 6):1625--1627, Jun 2016.

\bibitem{Kuhlmann2018}
L.~Kuhlmann, K.~Lehnertz, M.~P. Richardson, B.~Schelter, and H.~P. Zaveri.
\newblock {{S}eizure prediction - ready for a new era}.
\newblock {\em Nat Rev Neurol}, 14(10):618--630, Oct 2018.

\bibitem{Cook2013}
M.~J. Cook, T.~J. O'Brien, S.~F. Berkovic, M.~Murphy, A.~Morokoff, G.~Fabinyi,
  W.~D'Souza, R.~Yerra, J.~Archer, L.~Litewka, S.~Hosking, P.~Lightfoot,
  V.~Ruedebusch, W.~D. Sheffield, D.~Snyder, K.~Leyde, and D.~Himes.
\newblock {{P}rediction of seizure likelihood with a long-term, implanted
  seizure advisory system in patients with drug-resistant epilepsy: a
  first-in-man study}.
\newblock {\em Lancet Neurol}, 12(6):563--571, Jun 2013.

\bibitem{Lehnertz2016}
K.~Lehnertz, H.~Dickten, S.~Porz, C.~Helmstaedter, and C.~E. Elger.
\newblock {{P}redictability of uncontrollable multifocal seizures - towards new
  treatment options}.
\newblock {\em Sci Rep}, 6:24584, Apr 2016.

\bibitem{Fujiwara2016}
K.~Fujiwara, M.~Miyajima, T.~Yamakawa, E.~Abe, Y.~Suzuki, Y.~Sawada, M.~Kano,
  T.~Maehara, K.~Ohta, T.~Sasai-Sakuma, T.~Sasano, M.~Matsuura, and
  E.~Matsushima.
\newblock {{E}pileptic {S}eizure {P}rediction {B}ased on {M}ultivariate
  {S}tatistical {P}rocess {C}ontrol of {H}eart {R}ate {V}ariability
  {F}eatures}.
\newblock {\em IEEE Trans Biomed Eng}, 63(6):1321--1332, 06 2016.

\bibitem{Kerem2005}
D.~H. Kerem and A.~B. Geva.
\newblock {{F}orecasting epilepsy from the heart rate signal}.
\newblock {\em Med Biol Eng Comput}, 43(2):230--239, Mar 2005.

\bibitem{SchuBo2010}
A.~Schulze-Bonhage, F.~Sales, K.~Wagner, R.~Teotonio, A.~Carius, A.~Schelle,
  and M.~Ihle.
\newblock {{V}iews of patients with epilepsy on seizure prediction devices}.
\newblock {\em Epilepsy Behav}, 18(4):388--396, Aug 2010.

\bibitem{Poh2010}
M.~Z. Poh, T.~Loddenkemper, N.~C. Swenson, S.~Goyal, J.~R. Madsen, and R.~W.
  Picard.
\newblock {{C}ontinuous monitoring of electrodermal activity during epileptic
  seizures using a wearable sensor}.
\newblock {\em Conf Proc IEEE Eng Med Biol Soc}, 2010:4415--4418, 2010.

\bibitem{Poh2012}
M.~Z. Poh, T.~Loddenkemper, C.~Reinsberger, N.~C. Swenson, S.~Goyal, M.~C.
  Sabtala, J.~R. Madsen, and R.~W. Picard.
\newblock {{C}onvulsive seizure detection using a wrist-worn electrodermal
  activity and accelerometry biosensor}.
\newblock {\em Epilepsia}, 53(5):e93--97, May 2012.

\bibitem{Schmidhuber2015}
J.~Schmidhuber.
\newblock {{D}eep learning in neural networks: an overview}.
\newblock {\em Neural Netw}, 61:85--117, Jan 2015.

\bibitem{Truong2018}
N.~D. Truong, A.~D. Nguyen, L.~Kuhlmann, M.~R. Bonyadi, J.~Yang, S.~Ippolito,
  and O.~Kavehei.
\newblock {{C}onvolutional neural networks for seizure prediction using
  intracranial and scalp electroencephalogram}.
\newblock {\em Neural Netw}, 105:104--111, Sep 2018.

\bibitem{Kiral-Kornek2018}
I.~Kiral-Kornek, S.~Roy, E.~Nurse, B.~Mashford, P.~Karoly, T.~Carroll,
  D.~Payne, S.~Saha, S.~Baldassano, T.~O'Brien, D.~Grayden, M.~Cook,
  D.~Freestone, and S.~Harrer.
\newblock {{E}pileptic {S}eizure {P}rediction {U}sing {B}ig {D}ata and {D}eep
  {L}earning: {T}oward a {M}obile {S}ystem}.
\newblock {\em EBioMedicine}, 27:103--111, Jan 2018.

\bibitem{Chawla2005}
Nitesh~V. Chawla.
\newblock {\em Data Mining for Imbalanced Datasets: An Overview}, pages
  853--867.
\newblock Springer US, Boston, MA, 2005.

\bibitem{Jeni2013}
L.~A. Jeni, J.~F. Cohn, and F.~De~La~Torre.
\newblock {{F}acing {I}mbalanced {D}ata {R}ecommendations for the {U}se of
  {P}erformance {M}etrics}.
\newblock {\em Int Conf Affect Comput Intell Interact Workshops},
  2013:245--251, 2013.

\bibitem{Baud2018}
M.~O. Baud and V.~R. Rao.
\newblock {{G}auging seizure risk}.
\newblock {\em Neurology}, 91(21):967--973, Nov 2018.

\bibitem{VanBuren1958}
J.~M. VAN~BUREN.
\newblock {{S}ome autonomic concomitants of ictal automatism; a study of
  temporal lobe attacks}.
\newblock {\em Brain}, 81(4):505--528, Dec 1958.

\bibitem{Marshall1983}
D.~W. Marshall, B.~F. Westmoreland, and F.~W. Sharbrough.
\newblock {{I}ctal tachycardia during temporal lobe seizures}.
\newblock {\em Mayo Clin. Proc.}, 58(7):443--446, Jul 1983.

\bibitem{Smith1989}
P.~E. Smith, S.~J. Howell, L.~Owen, and L.~D. Blumhardt.
\newblock {{P}rofiles of instant heart rate during partial seizures}.
\newblock {\em Electroencephalogr Clin Neurophysiol}, 72(3):207--217, Mar 1989.

\bibitem{Robinson1966}
B.~F. Robinson, S.~E. Epstein, G.~D. Beiser, and E.~Braunwald.
\newblock {{C}ontrol of heart rate by the autonomic nervous system. {S}tudies
  in man on the interrelation between baroreceptor mechanisms and exercise}.
\newblock {\em Circ. Res.}, 19(2):400--411, Aug 1966.

\bibitem{Meisel2019}
C.~Meisel and K.~Bailey.
\newblock {{D}eep learning identifies signal-dependent information about the
  preictal state: a comprehensive assessment across ECoG, EEG and EKG}.
\newblock {\em submitted}, 2019.

\bibitem{Brinkmann2016}
B.~H. Brinkmann, J.~Wagenaar, D.~Abbot, P.~Adkins, S.~C. Bosshard, M.~Chen,
  Q.~M. Tieng, J.~He, F.~J. Munoz-Almaraz, P.~Botella-Rocamora, J.~Pardo,
  F.~Zamora-Martinez, M.~Hills, W.~Wu, I.~Korshunova, W.~Cukierski, C.~Vite,
  E.~E. Patterson, B.~Litt, and G.~A. Worrell.
\newblock {{C}rowdsourcing reproducible seizure forecasting in human and canine
  epilepsy}.
\newblock {\em Brain}, 139(Pt 6):1713--1722, 06 2016.

\bibitem{Kuhlmann2018b}
L.~Kuhlmann, P.~Karoly, D.~R. Freestone, B.~H. Brinkmann, A.~Temko,
  A.~Barachant, F.~Li, G.~Titericz, B.~W. Lang, D.~Lavery, K.~Roman,
  D.~Broadhead, S.~Dobson, G.~Jones, Q.~Tang, I.~Ivanenko, O.~Panichev,
  T.~Proix, M.~Nahlik, D.~B. Grunberg, C.~Reuben, G.~Worrell, B.~Litt, D.~T.~J.
  Liley, D.~B. Grayden, and M.~J. Cook.
\newblock {{E}pilepsyecosystem.org: crowd-sourcing reproducible seizure
  prediction with long-term human intracranial {E}{E}{G}}.
\newblock {\em Brain}, 141(9):2619--2630, Sep 2018.

\end{thebibliography}

\bibliographystyle{unsrt}


\section*{Acknowledgements}
This work was supported in part by the Epilepsy Research Fund (ERF). CM acknowledges support by a NARSAD Young Investigator Grant from the Brain \& Behavior Research Foundation. 

\section*{Ethical Publication Statement}
We confirm that we have read the Journal’s position on issues involved in ethical publication and affirm that this report is consistent with those guidelines.

\section*{Disclosure}
None of the authors has any conflict of interest to disclose.
TL received unrelated research support and device donations from Empatica, Inc.

\newpage

\newpage

\newpage
􏴎
\newpage

\newpage

\end{document}